\documentclass[prd,aps,showpacs,twocolumn,preprintnumbers,nofootinbib]{revtex4-1}
\usepackage[OT2,OT1]{fontenc}
\usepackage{amsmath,amssymb,amsfonts,amsthm}
\usepackage{graphicx}
\usepackage{hyperref}
\usepackage{bm}
\usepackage{color}

\def\be{\begin{equation}}
\def\ee{\end{equation}}
\def\ba{\begin{eqnarray}}
\def\ea{\end{eqnarray}}
\def\bs{\begin{subequations}}
\def\es{\end{subequations}}

\def\rme{e}
\def\rmd{d}

\def\p{\partial}
\def\bp{\bar\partial}
\def\cK{{\cal K}}

\def\cN{{\cal N}}
\def\cP{{\cal P}}
\def\cS{{\cal S}}
\def\cV{{\cal V}}

\def\N{\nabla}

\def\ds{d_{\rm S}}
\def\dh{d_{\rm H}}

\def\a{\alpha}
\def\b{\beta}
\def\de{\delta}
\def\la{\lambda}
\def\s{\sigma}

\def\g{\gamma}

\newcommand{\Eq}[1]{(\ref{#1})}
\def\com{\color{magenta}}
\def\cob{\color{blue}}

\newcommand{\arX}[1]{\href{http://arxiv.org/abs/#1}{{\ttfamily\com arXiv:#1}}}
\newcommand{\doin}[2]{\href{http://dx.doi.org/#1}{\cob #2}}

\begin{document}

\title{Diffusion in quantum geometry}
\author{Gianluca Calcagni}
\affiliation{Max Planck Institute for Gravitational Physics (Albert Einstein Institute),
Am M\"uhlenberg 1, D-14476 Golm, Germany}

\date{April 11, 2012}

\begin{abstract}
The change of the effective dimension of spacetime with the probed scale is a universal phenomenon shared by independent models of quantum gravity. Using tools of probability theory and multifractal geometry, we show how dimensional flow is controlled by a multiscale fractional diffusion equation, and physically interpreted as a composite stochastic process. The simplest example is a fractional telegraph process, describing quantum spacetimes with a spectral dimension equal to 2 in the ultraviolet and monotonically rising to 4 towards the infrared. The general profile of the spectral dimension of the recently introduced multifractional spaces is constructed for the first time.
\end{abstract}

\pacs{04.60.-m, 05.45.Df, 05.60.-k, 47.53.+n}

\preprint{AEI-2012-034}
\begin{center}
\preprint{\doin{10.1103/PhysRevD.86.044021}{Phys.\ Rev.\ D {\bf 86}, 044021 (2012)} \hspace{10cm} \arX{1204.2550}}
\end{center}

\maketitle


The spectral properties of effective quantum geometries show that the ultraviolet (UV) finiteness of independent theories of quantum gravity is universally associated with a lower spectral dimension of spacetime (typically, $\ds\sim2$) at small scales, while $\ds\sim 4$ in the infrared (IR). Instances are causal dynamical triangulations (CDT) \cite{cdt}, asymptotic safety (QEG) \cite{as,CES}, spin foams \cite{sf,COT}, noncommutative geometry \cite{nc}, Ho\v{r}ava-Lifshitz gravity \cite{hl}, and other approaches \cite{qs}. The change of dimension with the probed scale is known as dimensional reduction or dimensional flow \cite{caca}. Understanding its physical meaning is an important piece of the puzzle of quantum gravity, since multiscale behavior is deeply related to the renormalization properties of these theories. Differential geometry and ordinary calculus, as employed in general relativity and field theory, are inadequate to study this and other properties of quantum spacetimes, and stochastic processes and multifractal geometry can offer powerful tools of analysis and novel insight. While there is the tendency to label all multiscale spaces as ``fractal,'' the accumulated knowledge from these branches of mathematics permit to make sharper statements about the geometric and physical properties of quantum-gravity models. This philosophy inspired the revisiting of a recent problem, the construction of quantum field theories in fractal spacetimes, under a fresh perspective focused on an effective continuum geometry \cite{fr123}, in particular via the formalism of multifractional spacetimes \cite{frcs}.

After a sketch of the classical situation, we will argue that quantum geometry effectively modifies the diffusion equation. A critical appraisal of the latter will allow us to classify quantum geometries in terms of stochastic processes on one hand, and to get a precise back-up to the notion of ``fractal spacetime'' on the other hand. The aim is to reexamine the spectral dimension starting from its foundation and provide a general, model-independent and analytic description of dimensional flow, confirmed by quantum-gravity examples. This is possible thanks to the presence of universal features in the flow \cite{frc4}.

For a diffusion process to be meaningful, the solution $P$ of the transport equation must be nonnegative at all points. If $P$ is normalized to 1, it is interpreted as the probability to find the diffusing test particle at a given point. This probability distribution describes a stochastic process, i.e., a sequence or collection of random variables. We shall use the condition $P\geq 0$ as one of the guiding principles to identify the random process associated with a given behavior of quantum geometry. Here we do not pay attention to the techniques solving the diffusion equations; an expanded discussion is in \cite{frc4}. 


\emph{Classical spacetimes.} In a smooth classical spacetime with $D$ topological dimensions, the diffusion equation is
\be\label{de1}
\left(\p_\s-\N_x^2\right) P(x,x',\s)=0\,.
\ee
The parameter $\s\geq 0$ acts as an abstract ``time'' variable via the diffusion operator $\p_\s$, an ordinary first-order derivative. Writing $\s=\ell^2\bar\s$ in terms of a length scale $\ell$ and a dimensionless parameter $\bar\s$, Eq.\ \Eq{de1} is recast in the form $(\p_{\bar\s}-\ell^2\N_x^2)P=0$. The spatial generator $\N_x^2$ is the Laplacian in the given metric background in Euclidean signature. $x'$ is the initial point where diffusion starts. In translation-invariant spacetimes, the heat kernel $P$ depends on the difference $x-x'$, but in fractional spaces with nontrivial measure this is no longer true; therefore we keep the notation $P(x,x',\s)$ separate from the often-employed $u(x,\s)$ (fixing $x'=0$). Equation \Eq{de1} is not completely specified without the set of initial conditions at $\s=0$. The choice $P(x,x',0)=\delta(x-x')$ describes diffusion of a point particle starting at $x=x'$. Extended shapes of the probe are possible \cite{qs}, but the pointwise one allows us to explore the local structure of spacetime.

Ignoring curvature, the solution $P$ must be nonnegative for all $x$ and $x'$, and normalized as $\int\rmd^Dx P(x,x',\s)=1$. From the spatial trace of $P$, one gets the return probability $\cP(\s) :=(\int d^D x)^{-1}\int d^D xP(x,x,\s)$ and the spectral dimension
\be 
\ds := -2\frac{d\ln\cP(\s)}{d\ln\s}\,.\label{ds}
\ee
When divergent, the volume prefactor in the definition of $\cP$ can be regularized. In the case of a translation-invariant background, it cancels out with the position dependence in the numerator and $\cP(\s)=u(0,\s)$.

The normalized solution of Eq.\ \Eq{de1} is the Gaussian heat kernel $P(x,x',\s)=u_1(r,\s):=\rme^{-r^2/(4\s)}/(4\pi\s)^{D/2}$, 
where $r^2:=\sum_{\mu=1}^D|x_\mu-x_\mu'|^2$. Clearly, $P>0$. The return probability and spectral dimension read $\cP\propto \s^{-D/2}$ and $\ds=D$, respectively. There is no quantitative distinction between spectral and topological dimension. They are also equal to the Hausdorff dimension $\dh$ of spacetime, determining the scaling law of the volume of a $D$-ball of radius $R$, $\cV^{(D)}\propto R^{\dh}$. Notice that, because $\ell$ is the only scale, it is not possible to define a hierarchy of scales and the geometry (and $\ds$) is scale independent.

The ordinary diffusion equation \Eq{de1} is associated with a Wiener process $B(\s)$, the standard Brownian motion. $B$ is continuous in $\s$ with probability 1, $B(0)=x'$, and the increments of $B$ are governed by the Gaussian distribution $u_1$, $B(\s)-B(\s')\sim u_1(0,\s-\s')$ for $\s'<\s$.


\emph{Quantum geometry with fixed dimension.} Quantum geometry can emerge either by definition of a nonstandard texture of spacetime \cite{nc,hl,qs,fr123,frcs} or from the quantization of gravity \cite{cdt,as}, or for both reasons. Since the spectral dimension $\ds$ becomes then anomalous, quantum geometry effectively modifies Eq.\ \Eq{de1}. Modifications affect either the initial condition $P(x,x',0)$ (by a change of the definition of ``point particle'' on a quantum manifold \cite{qs} or of the delta distribution as in multifractional spaces \cite{frcs}), or the operator $\p_\s$ as in QEG \cite{CES}, or the Laplacian $\N^2_x$ (via the change in the differential structure and/or the presence of one or more fundamental quantum scales $\ell_n$, such as the Planck scale or the label-dependent lengths of the simplices in a cellular complex), as in QEG \cite{as,CES}, Ho\v{r}ava gravity \cite{hl}, multifractional theory \cite{frcs}, and CDT, spin foams and simplicial gravity in general \cite{cdt,sf,COT}.

Dimensional flow is still inadequately understood and a classification of the possible diffusion equations should help to control the physics of the above-mentioned (as well as other) models of quantum gravity. It is instructive to specialize first to the case of fixed dimensionality (no scale hierarchy). We concentrate on the continuum formulation of fractional calculus, which guarantees anomalous (in particular, fractal) geometric properties of spacetimes \cite{frcs,frc4} and anomalous correlations in diffusion problems (e.g., \cite{diff}). Multifractional theory is a model of quantum gravity in its own right (like Ho\v{r}ava gravity, it is a traditional perturbative field theory but built on an anomalous spacetime), although it can serve as a framework to understand other proposals \cite{CES,nc}. We ignore curvature. The latter modifies the spectral properties of spacetime even in a classical setting, except in the UV limit $\s\to 0$. Quantum effects, however, often modify spacetime globally even in the absence of curvature, which motivates the assumption (see also \cite{CES,caca}). We replace $\p_x^2$ with the operator
\be\label{K}
\cK_{\g,\a} := -\frac{1}{\sqrt{v_\a(x)}}\,\frac{\sum_\mu({}_\infty\p^{2\g}_\mu+{}_\infty\bp^{2\g}_\mu)}{2\cos(\pi\g)}\left[\sqrt{v_\a(x)}\,\cdot\right],
\ee
where $v_\a(x)\propto \prod_\mu|x_\mu|^{\a-1}$ is the measure weight of the ambient space (the singularity in $x^\mu=0$ is integrable and does not pose particular problems for the classical and quantum dynamics), $0<\a\leq 1$ and $0<\g\leq m$ are real parameters, and we make use of left and right Liouville-Caputo fractional derivatives: for each direction, $({}_\infty\p^{2\g} f)(x) \propto\int_{-\infty}^{x} \rmd x'(x-x')^{m-1-2\g}\p^m_{x'}f(x')$, $({}_\infty\bp^\g f)(x) \propto (-1)^m\int_{x}^{\infty} \rmd x'(x'-x)^{m-1-2\g}\p^m_{x'}f(x')$.
When $2\g=m$ is integer, ${}_\infty\p^{m}=(-1)^m{}_\infty\bp^{m}=\p^m$. Definition \Eq{K} is such that, in a suitable domain, the operator $\cK_{\g,\a}$ is self-adjoint \cite{frcs,frc4}.

We classify the stochastic and geometric properties associated with the diffusion equation
\be\label{de2}
\left(\p_\s-\cK_{\g,\a}\right)P=0\,,
\ee
with initial condition $P|_{\s=0}=[v_\a(x)v_\a(x')]^{-1/2}\delta(x-x')$. One can show that $\ds=D\a/\g$ \cite{frc4}.
\begin{enumerate}
\item[(i)] When $\g=1=\a$, we recover ordinary diffusion and $\ds=D$. For $\a\neq 1$, this is ordinary Brownian motion but on a fractal spacetime with $\ds=D\a$.
\item[(ii)] For $0<\g<1$ and $\a=1$, we have a superdiffusive L\'evy process with $\ds=D/\g>D=\dh$. This does \emph{not} correspond to a fractal spacetime ($\ds\leq \dh$ for fractals). For $0<\g,\a<1$, one has a L\'evy process on an anomalous spacetime (fractal if $\a\leq\g$).
\item[(iii)] When $\g>1$, the solution of \Eq{de2} is no longer nonnegative definite and the equation must be modified. In fact, one can include a source term $\cS(x,x',\s)$, which does not alter the spectral dimension.
 Hence, overlooking the check that $P\geq 0$ for the \emph{Ansatz} \Eq{de2} might result in the correct spectral dimension but a wrong diffusion equation.
\end{enumerate}
Processes associated with nonhomogeneous equations may be non-Markovian (i.e., future states depend also on past states) even if they are meaningful in a probabilistic sense. An example is the quartic equation
\be\label{mitF}
\left(\p_\s-\N^4_x\right)u(x,\s)=(\pi\s)^{-1/2}\N^2_x u(x,0)\,,
\ee
with source given by the initial condition. The solution gives the same $\ds=D/2$ as the naive Eq.\ \Eq{de2} with $\g=2$, but the presence of the source guarantees that $u\geq0$. Equation \Eq{mitF} governs an \emph{iterated Brownian motion} (IBM) \cite{ibm}. Given two independent Wiener processes $B_{1,2}$, IBM is defined as $X_{\rm IBM}(\s):=B_1[|B_2(\s)|]$, where $B_2$ acts as a clock to $B_1$. In general, there exists a deep connection between higher-order diffusion equations with integer time, iterated stochastic processes, and diffusion equations with fractional time $(\p_\s^\b-\N^2_x)u=0$, with $0<\b\leq 1$ and $\p^\b_\s$ the left Caputo derivative (lower terminal $\s'=-\infty$ replaced by $\s'=0$). The solution $u$ is positive definite \cite{OB2}. The process described by this equation (\emph{fractional Brownian motion}) is subdiffusive and on average it takes longer (with respect to Brownian motion) for the particle to cover a certain distance. Equation \Eq{mitF} can be regarded as the ``iteration'' of the fractional equation with $\b=1/2$, with the \emph{same} solution $u$.

Extending the discussion to a nontrivial spacetime measure, the spectral dimension associated with the fractional diffusion equation $(\p_\s^\b-\cK_{\g,\a})P=0$ is $\ds=(\b/\g)\dh$, where $\dh=D\a$ \cite{frcs,frc4}. A fractal configuration is obtained whenever $\b\leq\g$.


\emph{Multiscale quantum spacetimes.} We now make a conceptual step of relevance for quantum gravity. In complex systems, the adoption of a diffusion equation is motivated by phenomenology. Given a set of experiments evidencing some anomalous scaling laws, one proposes an ad-hoc diffusion equation reproducing those scalings. The theory is then further tested against experiments. Or else, one defines the stochastic process underlying a certain physical system, and from its probability distribution one infers the correct diffusion equation. For instance, IBM provides a stochastic description of diffusion of a particle trapped in a crack \cite{crack}, the latter being modeled by a random fractal set whose pattern is the graph of a Brownian motion. In quantum gravity, on the other hand, we do not have experiments but fragmentary knowledge such as the existence of anomalous scaling behaviors in the UV. This information determines the differential order of the operators in the diffusion equation, i.e., the number and values of the plateaux in the $\ds$ profile, \emph{but it may be unable to fix the diffusion equation univocally}. In particular, the monotonic slopes between the plateaux can vary from model to model of the same physical system, but they may be not falsifiable features; their nonuniqueness can be usually traced back to details of the theory such as regularization schemes \cite{CES,frc4}. Also, dimensional flow may be insensitive of the presence of source terms [Eq.\ \Eq{mitF} without source or with flipped sign in front of $\N^4$ would still give the same $\ds$], and we must resort to positivity of the solution $P$ to fix more details of the diffusion equation. In turn, once we determine a reasonable diffusion equation with probabilistic interpretation, we can also find the stochastic process associated with that, thus physically characterizing quantum geometry.

Without further input from the theory except the UV and IR behaviors, we can reproduce the \emph{whole} dimensional flow by applying the techniques of multiscale phenomena and multifractal geometry to the texture of spacetime itself \cite{frc4}. The generalization of the diffusion equation (with solution $P\geq 0$) to a multiscale process is realized by summing over all possible values of $\a$, $\b$, $\g$:
\be\label{muf1}
\sum\nolimits_n\left(\xi_n\p_\s^{\b_n}-\zeta_n\cK_{\g_n,\a_n}\right)\,P(x,x',\s)=\cS(x,x',\s)\,,
\ee
where $\xi_n$ and $\zeta_n$ are dimensionful couplings which depend on the characteristic scales of the system. Typically, there is only a finite number $N$ of terms in physical systems, so the sum representation \Eq{muf1} is realistic. The number $N-1$ of characteristic scales (hidden in $\xi$ and $\zeta$) determines the number $N$ of plateaux (asymptotic regimes) in the profile of $\ds$. A multiscale phenomenon is always defined by the relative size of the scales, not by an absolute hierarchy. This means that we can choose any of the $N$ scales $\ell_n$ to represent the scale $\ell$ probed by a measurement. If we order the scales of the system as $\ell_1<\ell_2<\dots<\ell_N$, we can take the largest as $\ell=\ell_N$. Thus, there are $N-1$ (not $N$) scales with the physical meaning of characteristic lengths. The spectral dimension is fixed when $N=1$; for $N=2$, it has two asymptotic values $\ds\sim {\ds}_{1,2}$ in the regimes $\ell\ll\ell_1$ and $\ell\gg\ell_1$, with a monotonic transient phase in between; for $N=3$, there will be also an intermediate plateau where $\ds\sim {\ds}_3$; and so on.

The first example is an interaction of Gaussian and anomalous dynamics which can describe certain turbulent media \cite{WeZ1}. The diffusion equation is $(\p_\s-\p_x^2-\zeta_1\cK_{\g,1})u=0$, $u(x,0)=\de(x)$, 
 where $0<\g<1$ and we write the constant $\zeta_1=\ell_1^{-2(1-\g)}$ in terms of a characteristic length. As the analytic solution shows, the transport is of L\'evy type at large scales $\ell=k^{-1}\gg \ell_1$ ($\ds\sim 1/\g>1=\dh$) and normal at small scales $\ell\ll \ell_1$ ($\ds\sim 1$). From the perspective of quantum spacetimes, this model is multiscale but not multifractal. Profiles of $\ds$ overshooting the Hausdorff and topological dimensions appear also in lattice \cite{COT} and noncommutative geometries (last reference in \cite{nc}), with some caveats \cite{frc4}.

A second example is a fractional diffusion equation with two diffusion operators $\p^{\b_{1,2}}_\s$. To see its neat stochastic interpretation, we recall some results on the so-called telegraph processes \cite{OB1}. A telegraph process is defined as $V(\s)=V(0)\,(-1)^{\cN(\s)}$, where $V(\s)$ is the velocity of a particle at time $\s$ running on the real line, $V(0)$ is the initial velocity which is $\pm c$ with equal probability, and $\cN$ is the cumulative number of events of a homogeneous Poisson process with rate $\la>0$. The velocity of the particle flips direction back and forth, hence the name ``telegraph.'' The position of the particle at time $\s$ is the integrated telegraph process $T(\s)=V(0)\int_0^s\rmd s\,(-1)^{\cN(s)}$. We consider a composite process called Brownian-time telegraph process or \emph{fractional telegraph process}, $X_{{\rm FTP}}(\s):= T[|B(\s)|]$. This motion is governed by the diffusion equation $(\p_\s+2\la\p_\s^{1/2}-c^2\p_x^2)u=0$. The solution of the fractional telegraph equation and its generalizations is nonnegative and unique \cite{OB1}. In the double limit $\la,c\to+\infty$, $\la/c^2\to {\rm const}$, the stochastic process reduces to an IBM. Recasting these results in the language of multifractal spacetimes and extending to $D$ dimensions, we set $[\s]=0$, $c=\ell_*^2$ as the characteristic scale, and $\ell=\ell_*/(2\la)$ as the probed scale. In the limit $\ell\gg\ell_*$, diffusion is Gaussian (Brownian process, $\ds\sim D$). At small scales $\ell\ll\ell_*$, one reaches a regime where diffusion is fractional and given by an IBM ($\ds\sim D/2$). In between, diffusion in quantum spacetime obeys the law of a fractional telegraph process.

The monotonic profile $\ds(\ell)$ of this single-scale spacetime can be plotted from the analytic form of the return probability \cite{frc4}. The probability distribution for more complicated multiscale spacetimes can be computed as well, but here we show how all these profiles are easily reproduced in the framework of multifractional geometry ($\a_n\neq1$) when $\b=1=\g$ and $\cS=0$. We argue that the coefficients $\zeta_n$ in \Eq{muf1} have the natural form $\zeta_N=1$ and
\be\label{zeta}
\zeta_1(\ell)=(\ell_1/\ell)^{2}\,,\qquad \zeta_n(\ell)=[\ell_n/(\ell-\ell_{n-1})]^{2}\,,
\ee
where $\ell=\ell_N$. First, we notice that the Laplacians all have the same order $2$, so the coefficients $\zeta_n$ all have the same scaling dimension and, in particular, we can always make them dimensionless. Write $\zeta_n$ as the ratio of some length scales, $\zeta_n=(l_{A,n}/l_{B,n})^q$. Without loss of generality, one can choose $q=2$ so that the spatial generator of the diffusion equation can be rendered dimensionless, in the form $\sum_n(l_{A,n})^{2}\cK_{1,\a_n}$. Now, the $n$th term dominates over the others at scales $\ell\ll\ell_n$, while at scales smaller than $\ell_{n-1}$ the $(n-1)$th term takes the lead, so the smallest possible scale $\ell$ at which the $n$th term dominates is $\ell\sim\ell_{n-1}$. Thus, we set $l_{A,n}=\ell_n$ and $l_{B,n}=\ell-\ell_{n-1}$. In other words, the dimensional flow is always measured starting from the lowest of two scales $\ell_{n-1}$ to the next $\ell_n$, and relatively to the latter, which sets a gauge for the rods. Since $\ell=\ell_N$ is the probed scale, $\zeta_N\equiv1$.
\begin{figure}[t]
\includegraphics[width=8.3cm]{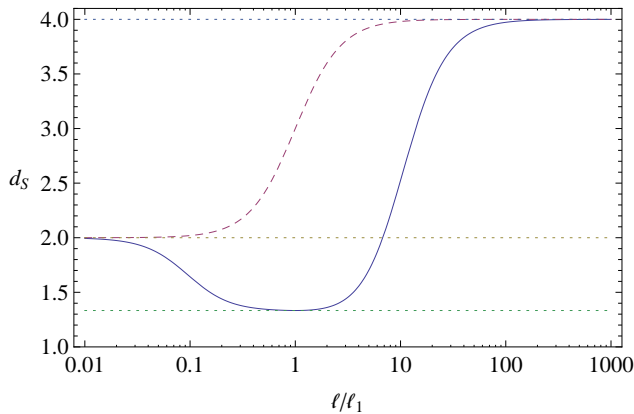}
\caption{\label{fig1} The spectral dimension $\ds(\ell)$ in $D=4$ for a multifractional model and normal diffusion ($\b=1=\g$, $\ds=\dh$) with single scale (dashed curve) and two scales (solid curve).}
\end{figure}

We can plot the spectral dimension for any given profile $\a(\ell)$. Upgrading on \cite{frcs}, we motivate a realistic profile $\a(\ell)$ as an approximation of the sum in \Eq{muf1}. Consider first the $N=2$ case with $\a_1\neq 1$ and $\a_2=1$. In one dimension, and by Eq.\ \Eq{K}, $(\p_x^2+\zeta_1\cK_{1,\a_1}^{\rm E})P = \{(1+\zeta_1)\cK_{1,\a_1(\ell)}^{\rm E}+\zeta_1(1-\a_1)^2/[4(1+\zeta_1)x^2]\}P$, where $\a_1(\ell):=[1+\zeta_1(\ell)\,\a_1]/[1+\zeta_1(\ell)]$. For both small and large $\zeta_1$ the kinetic term in this expression dominates over the potential term, so the profile $\a_1(\ell)$ defines an effective fractional charge throughout the dimensional flow. With $N$ coefficients $\a_n$, $\a_N=1$, the effective charge reads
\be\label{profN}
\a_{N-1}(\ell):=\frac{1+\sum_{n=1}^{N-1}\zeta_n(\ell)\,\a_n}{1+\sum_{n=1}^{N-1}\zeta_n(\ell)}\,.
\ee

For two entries ($N=2$, $\a_1=2/D$, $\a_2=1$, one scale), dimensional flow is such that $\ds\sim D$ in the IR and $\ds\sim D\a_1=2$ in the UV, with no intermediate regime in between. This is the type of flow considered in \cite{fr123,frcs} and is shown in Fig.\ \ref{fig1} (dashed curve), in quantitative agreement with the fractional-telegraph profile which is matched by tuning the scale length \cite{CES}. A two-scale profile $\ds(\ell)=4\a_2(\ell)$ with $\a_1=1/2$, $\a_2=1/3$ and $\ell_2=10\ell_1$ is also plotted (solid curve). At $\ell=0$, $\ds=2$. At $\ell\sim \ell_1$, the spectral dimension acquires the minimum value $\ds=4/3$. At scales $\ell\leq\ell_2$, the diffusion process corresponds to a recurrent random walk, where $\ds<2$. Well above the larger critical scale, $\ell\gg\ell_2$, $\ds$ hits the IR value $\sim 4$. Notably, the profiles in Fig.\ \ref{fig1} reproduce the dimensional flow of asymptotically-safe quantum gravity in two different realizations, without or with matter. This is consistent both with the fact that the single-scale curve of QEG does come from a fractional telegraph process equation and with a reinterpretation of the renormalization group flow in terms of measurements in multifractional geometry \cite{CES}.


\end{document}